\documentclass[fleqn,twoside]{article}
\usepackage{espcrc2}

\usepackage{graphicx}
\usepackage[figuresright]{rotating}

\newcommand{\AmS}{{\protect\the\textfont2
  A\kern-.1667em\lower.5ex\hbox{M}\kern-.125emS}}

\title{The propagation of ultra-high energy tau leptons and neutrinos while skimming the Earth}

\author{O. Blanch Bigas\address[LPNHE]{LPNHE, CNRS/IN2P3 \& Universit\'e Paris 6-7, Paris, France}, O. Deligny\address[IPN]{IPN, CNRS/IN2P3 \& Universit\'e Paris Sud, Orsay, France}, K. Payet\address[LPSC]{Universit\'e Joseph Fourier Grenoble 1, CNRS/IN2P3, INPG, Grenoble, France}, V. Van Elewyck\addressmark[IPN] \address{AstroParticule et Cosmologie (UMR 7165) \& Universit\'e Paris 7, Paris, France}}
       
\begin{document}

\begin{abstract}

\vspace{1pc}
\end{abstract}

\maketitle

\section{INTRODUCTION}

With the advent of a new generation of large-scale detectors of cosmic radiation, the observation of high-energy neutrinos produced in distant astrophysical sites (or possibly by other, more exotic, mechanisms) has become one of the major challenges of astroparticle physics\cite{nurev}. Given the large distances traveled, approximately equal fluxes in $\nu_e , \nu_\mu$ and $\nu_\tau$ are expected on Earth as a result of flavour mixing and oscillations~\cite{athar,halzen}.

The detection of Earth-skimming  $\nu_\tau$'s (through the observation of
the shower induced by the emerging $\tau$ lepton\cite{nutau}) has become a very
promising strategy for the observation of ultra-high energy (UHE) cosmic neutrinos. This
method has been used recently by cosmic-ray experiments such as HiRes and
the Pierre Auger Observatory to put competitive limits on the flux of cosmic
neutrinos in the energy range  $10^{-1}\ \mathrm{EeV} \leq E_\nu \leq 10^{2}\ \mathrm{EeV}$ \cite{augernu,hiresnu}.

The sensitivity to such UHE Earth-skimming neutrinos crucially depends on the parameters of the $\nu_\tau$ and $\tau$ propagation through the terrestrial crust, and on the correct estimation of the flux of $\tau$'s that emerge from the Earth (see Fig. \ref{fig:simul}). This problem has been widely discussed in different contexts and with different approximations (see e.g. the references provided in \cite{nousEloss,nousreg}). An exhaustive treatment should account for $\tau$ and $\nu_\tau$ neutral- and charged-current interactions with nucleons, $\tau$ decay and energy losses. 

However the full coupled transport equations admit no analytical solution, and they are usually treated numerically (or semi-analytically) in a simplified framework where several effects are neglected, such as the possibility of multiple regenerations of the $\nu_\tau$, the weak interactions of the $\tau$, and the stochastic nature of its energy losses. Significant uncertainties in the calculation also derive from the poor knowledge of the relevant cross-sections at UHE, where no direct measurements exist. This is particularly true for the neutrino-nucleon interactions and for the photonuclear contribution to the tau energy losses, which both rely on extrapolations of the structure functions beyond the range of $(x,Q^2)$ probed at accelerators and which could be affected by the onset of new physics beyond the Standard Model.

These considerations demonstrate the need to assess the impact  of such simplifications on the determination of the final flux of emerging $\tau$'s. We present here such a study of the propagation in standard rock of tau leptons and neutrinos with both mono-energetic and power-law spectra. 

The common framework and the strategy used to solve the transport equations is briefly described in Section \ref{sec:tpt}. In Section \ref{sec:Taulosses} we focus on the $\tau$ energy losses and compare the results obtained with a full stochastic treatment and within the continuous approximation. Then, we investigate in Section \ref{sec:reg} the impact of the $\nu_\tau \rightarrow \tau \rightarrow \nu_\tau$ regeneration chain both in the standard case and in non-standard scenarios for the neutrino-nucleon
interactions and for the tau energy losses. The conclusions are presented in Section \ref{sec:conc}, where we also discuss the impact of our studies on the sensitivity of current extensive air shower detectors to Earth-skimming $\nu_\tau$'s. 

\section{$\nu_\tau$ AND $\tau$ PROPAGATION: GENERAL TRANSPORT EQUATIONS}
\label{sec:tpt}
\begin{figure}[!b]
    \centering
    \vspace*{-0.5cm}
    \includegraphics[bbllx=1,bburx=548,bblly=350,bbury=490,width=7.5cm,clip=]{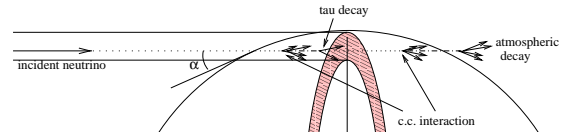}
    \vspace*{-0.8cm}
    \caption{\small{Geometry of the transport problem.}}
    \label{fig:simul}
\end{figure}

The geometry of the propagation problem is described in
Fig.\ref{fig:simul}, where an example of regeneration chain through multiple CC interactions and $\tau$ decays is 
sketched. Given a beam of parallel neutrinos incident on the Earth, the problem becomes uni-dimensional. The 
flux of the tau leptons that emerge  only depends on the amount of crossed rock which, in the hypothesis of a spherical 
Earth with a constant density in its crust, is given by the incident nadir angle $\alpha$, 
which is also the angle of the emerging tau.

Two coupled, integro-differential equations describe the evolution of the  $\nu_\tau$ and $\tau$ fluxes, $\Phi_{\nu_\tau}$ and $\Phi_{\tau}$, along their paths
through the Earth, accounting for all possible production and
absorption processes taking place within an infinitesimal displacement $dx$: 
%\vspace*{-0.5cm}
\begin{eqnarray}
    &&\hspace*{-0.5cm}\frac{\partial \Phi_{\nu_\tau}(E,x)}{\partial x} =
    - \frac{\Phi_{\nu_\tau}(E,x)}{\lambda_{\nu_\tau}^{CC}(E)}
    -\frac{\Phi_{\nu_\tau}(E,x)}{\lambda_{\nu_\tau}^{NC}(E)} \nonumber\\
    &&+ \rho\,\mathcal{N}_A\,\int\frac{\mathrm{d}y}{1-y}\Phi_{\nu_\tau}
    \bigg(\frac{E}{1-y},x\bigg)
    \frac{\mathrm{d}\sigma_{\nu_\tau}^{NC}(y,\frac{E}{1-y})}{\mathrm{d}y}\nonumber\\
    &&+ \rho\,\mathcal{N}_A\int\frac{\mathrm{d}y}{1-y}
    \Phi_{\tau}\bigg(\frac{E}{1-y},x\bigg)
    \frac{\mathrm{d}\sigma_{\tau}^{CC}(y,\frac{E}{1-y})}{\mathrm{d}y}\nonumber\\
    &&+ \frac{1}{c}\int\frac{\mathrm{d}y}{1-y}
    \Phi_{\tau}\bigg(\frac{E}{1-y},x\bigg)
    \frac{\mathrm{d}\Gamma_\tau (y,\frac{E}{1-y})}{\mathrm{d}y}
    \label{eq:PhiNuTau}
\end{eqnarray}

where $\lambda_{\nu_\tau}^{NC}$ and $\lambda_{\nu_\tau}^{CC}$ are the
mean free paths corresponding respectively to NC and
CC interactions of the incident $\nu_\tau$, while
$\sigma_\tau^{CC}$ corresponds to the $\tau$ CC
interaction, which regenerates a $\nu_\tau$. $\Gamma_\tau$ is the
tau lepton lifetime. Similarly, the equation for the $\tau$ reads:
%\vspace*{-0.5cm}
\begin{eqnarray}
\vspace*{-0.5cm}
  &&\hspace*{-1cm}\frac{\partial \Phi_\tau(E,x)}{\partial x} =
  -\frac{\Phi_\tau(E,x)}{\lambda_\tau^{dec}(E)}  -\frac{\Phi_{\tau}(E,x)}{\lambda_\tau^{CC}(E)} \nonumber\\
&&\hspace*{-1cm}  -\frac{\Phi_{\tau}(E,x)}{\lambda_\tau^{NC}(E)}
  -\sum_i \bigg[
  \frac{\Phi_\tau(E,x)}{\lambda^i_\tau(E)}
  \nonumber\\
  \hspace*{-1cm}&-&\frac{\rho \mathcal{N}}{A}
  \int\frac{\mathrm{d}y}{1-y}\Phi
  \bigg(\frac{E}{1-y},x\bigg)
  \frac{\mathrm{d}\sigma^i_{\tau}}
  {\mathrm{d}y}\bigg(y,\frac{E}{1-y}\bigg)
  \bigg]\nonumber \\  
&&\hspace*{-1cm}+\rho\,\mathcal{N}_A\,\int\frac{\mathrm{d}y}{1-y}
    \Phi_{\tau}\bigg(\frac{E}{1-y},x\bigg)
    \frac{\mathrm{d}\sigma_\tau^{NC}(y,\frac{E}{1-y})}{\mathrm{d}y}\nonumber\\
    &&\hspace*{-1cm}+\rho\,\mathcal{N}_A\int\frac{\mathrm{d}y}{1-y}
    \Phi_{\nu_\tau}\bigg(\frac{E}{1-y},x\bigg)
    \frac{\mathrm{d}\sigma_{\nu_\tau}^{CC}(y,\frac{E}{1-y})}{\mathrm{d}y}
  \label{eq:PhiTau1}
\end{eqnarray}

where the index $i$ runs over the $\tau$ energy loss processes, namely pair production, bremsstrahlung and photonuclear interactions.

At the energies of interest for this study, the relevant
electromagnetic processes that the tau lepton undergoes
are the bremsstrahlung, the pair production and the photonuclear
interactions. The dominant source of $\tau$ energy losses at $E_\tau > 6$ EeV comes from the photonuclear interactions, whose description relies on the proper modelisation
of the nucleon structure functions at low $x$ and high $Q^2$. Several calculations of the corresponding cross section, based on different theoretical models, can be found in the literature\cite{stdelosses,dutta1,PT,parente}. All of them give comparable results except for \cite{PT} which gives a significantly higher loss rate, and for \cite{parente} which gives a rate more than a factor of 2 lower than the standard one at the highest energies ($\sim 10^{12}$ GeV). 

As for the neutrino-nucleon cross-section, a benchmark parameterization based on recent HERA data is given in \cite{sarkar}. The same paper points out another, more speculative, approach based on the color glass condensate formalism\cite{CGC} which gives a much lower cross-section. On the other hand, plenty of models using new physics predict instead an enhancement of $\sigma_{\nu N}$ at UHE\cite{han}.     

\section{TAU ENERGY LOSSES: CONTINUOUS VS STOCHASTIC APPROACH}
\label{sec:Taulosses}

In order to investigate whether the stochasticity of this process has an effect on the description of the propagation of UHE $\tau$'s through the Earth, we will first work in a slightly  simplified framework where we neglect the $\tau$ weak interactions (a reasonable assumption at UHE) as well as the possibility of $\tau \rightarrow \nu_\tau \rightarrow \tau$ regeneration chains (an effect which will be studied in Section \ref{sec:reg}). For the  $\tau$ photonuclear interactions, we use here the standard parameterization of the PDFs given in \cite{dutta1} based on Regge theory.

When the differential cross-sections exhibit a peak near
$y$=0, as it is indeed the case for the processes that govern the $\tau$ energy losses, the integrals are dominated by the behavior of the integrands
around 0, in such a way that an expansion of these integrands can
be performed. At first order in $y$, and with the simplificating assumptions described hereabove, this gives the following transport equation for the $\tau$:
\begin{equation}
  \frac{\partial \Phi(E,x)}{\partial x} =
  -\frac{\Phi(E,x)}{\lambda_{dec}(E)}
  +\rho \frac{\partial}{\partial E}
  \bigg(E\beta(E)\Phi(E,x)\bigg)
  \label{eq:PhiTau2}
\end{equation}
where we have introduced the standard notation
\begin{eqnarray}
  \beta(E)=\frac{\mathcal{N}}{A}\sum_i
  \int_{y_{min}^i}^{y_{max}^i}\mathrm{d}y\,y\,
  \frac{\mathrm{d}\sigma_{i}}{\mathrm{d}y}(E,y).
  \label{eq:beta}
\end{eqnarray}

Within the approximation of continuous energy losses,
the average energy lost by the $\tau$ per unit distance $\mathrm{d}E_\tau/\mathrm{d}x$
is, at UHE, assumed to be proportional to the mean
inelasticity of each process which is directly related to $\beta(E)$, yielding the familiar expression :
\begin{eqnarray}
	\frac{\mathrm{d}E_\tau}{\mathrm{d}x} = -\rho \ \beta(E_\tau)\ E_\tau.
	\label{eq:dedx}
\end{eqnarray}
In that case, Eqn.~\ref{eq:PhiTau2} can
be easily integrated, leading to the following expression :
\begin{eqnarray}
	\Phi(E,x) &=& \Phi_0(\tilde{E}_0)\	 \times \nonumber\\
&& \hspace*{-1,5cm}\times	\exp \ \int_0^x\mathrm{d}u \bigg(\frac{\partial}{\partial E} \gamma(\tilde{E}_u)
	-\frac{1}{\lambda_{dec}(\tilde{E}_u)} \bigg) 
	\label{eq:sol}
\end{eqnarray}
where $\gamma(E)=\rho E\beta(E)$ and $\tilde{E}_{v}$ the solution of
\begin{eqnarray}
	\int_{\tilde{E}_{v}}^E \frac{\mathrm{d}E_\tau}{\gamma(E_\tau)}=v-x.
\end{eqnarray}
In the following, we use Eqn.~\ref{eq:sol} to compute
any propagated flux of tau leptons when referring to as
the continuous energy losses (CEL) approximation. 
 
What we want now is to compare the accuracy of results obtained by using this approximation with those from a fully simulated transport equation including the stochastic terms. Therefore we use a Monte-Carlo generator sampling all the interactions, which separates the losses into two components~\cite{music}:
a continuous one where the rate of the losses is large
($y\in[y_{min},y_{cut}]$), and a stochastic one where the
differential cross sections lead to more catastrophic losses
but with a weaker rate ($y\in[y_{cut},y_{max}]$). A good compromise to reproduce the stochastic features using a reasonably fast code is to take $y_{cut}=10^{-3}$. 
  
 \begin{figure}[!b]
 \vspace*{-1cm}
    \centering
    \includegraphics[width=7.5cm,height=6cm]{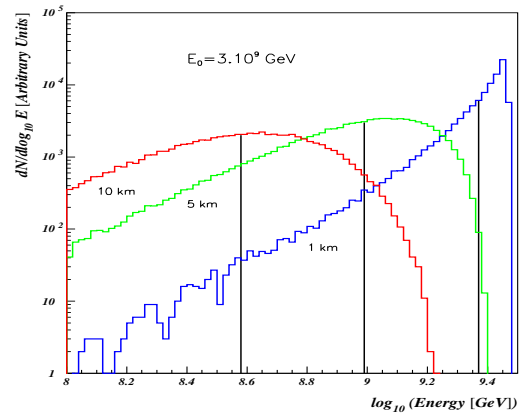}
\vspace*{-1cm}
    \caption{\footnotesize{Energy log$_{10}$-distributions of taus of $E_0=3\cdot 10^9$ GeV after propagation in standard rock for
    3 different depths (1, 5 and 10 km). Also shown are the mean
    values of the energy distributions.}}
    \label{fig:mono}
\end{figure}

\vspace*{0.3cm}
We show in Fig.~\ref{fig:mono} the simulated distributions of the energy of an incident monoenergetic tau beam after crossing different depths of standard rock, as obtained in the stochastic framework, together with the mean of the energy distributions, which are the values used within the CEL approximation. 

\begin{figure}[!b]
 \vspace*{-0.7cm}
    \centering
    \includegraphics[width=7.5cm,height=6cm]{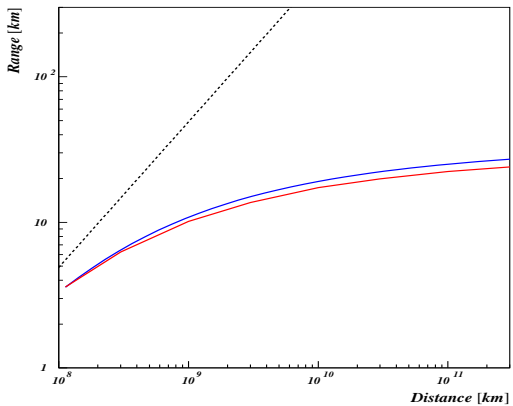}
\vspace*{-1cm}
    \caption{\footnotesize{Range  $R(E_0)=\int_0^\infty \mathrm{d}x \ P_{surv}(E_0,x)$ of the tau calculated with continuous (top curve) and stochastic (bottom curve) energy losses. For comparison, the decay length of the tau (dashed line) is also shown.}}
    \label{fig:range}
\end{figure}

After 1 km, many  of the simulated events still carry a large fraction of the initial energy $E_0$, meaning that these particles did not undergo many interactions. However, the distribution is asymmetric and there
is a large tail of events undergoing hard losses. For longer paths in rock, fluctuations in the energy losses increase, resulting in a broadening of the distribution and a smoothening of the high-energy cutoff. The stochasticity of $\tau$ energy losses thus globally leads to a smaller survival probability for the $\tau$ than in the CEL approximation; the difference in the $\tau$ range is however always small, as can be seen in Fig.\ref{fig:range}.

\vspace*{0.3cm}

\begin{figure}[!b]
 \vspace*{-0.7cm}
    \centering
    \includegraphics[width=7.5cm,height=6cm]{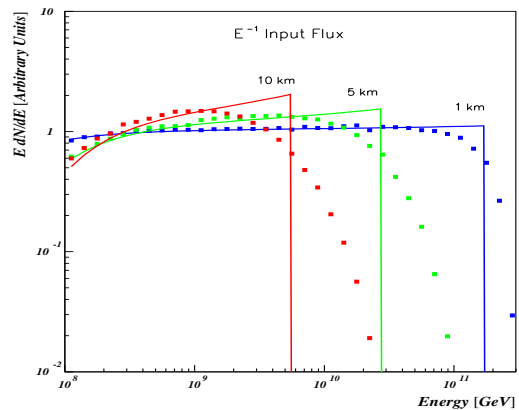}
\vspace*{-1cm}
    \caption{\footnotesize{Comparison of the emerging $\tau$ fluxes (normalized to the injection flux) obtained with the CEL approximation (Eqn.\ref{eq:sol}) and with the full Monte-Carlo calculation including the stochastic effects  (squared dots). An $E^{-1}$ flux
    between $10^8$ GeV and $3\cdot 10^{11}$ GeV is used at injection.
    Three distances of propagation in rock are shown: 1 km,
    5 km and 10 km.}}
    \label{fig:e1}
\end{figure}

We have also looked for  possible distorsions on the propagated fluxes
due to the stochastic effects in the case of continuous $E^{-1}$ and $E^{-2}$
tau injection spectra in the range $10^8$ GeV $\leq E_\tau \leq 3\cdot 10^{11}$ GeV. The cutoff for the
maximum energy is chosen to be sharp to exhibit most clearly
the different behaviors of the propagated fluxes. We compare the results obtained with the stochastic and the continuous energy losses in Fig.\ref{fig:e1} for the case of an $E^{-1}$ incident $\tau$ flux, where the differences are more visible.  

Here again, the sharp cutoff present in the CEL approximation gets smoothed when accouting for stochastic processes: there are indeed fluctuations affecting a small fraction of particles which undergo less interactions and less hard losses. The energy range on which this broadening occurs increases with the depth traversed by the tau, and with the hardness of the incident flux. It is the only important distorsion induced by
the stochastic processes. This means that, as far as we are dealing with continuous incident spectra behaving as power-laws,
there is a compensation between positive and negative fluctuations
of the energy losses everywhere in the considered energy range, except near
the high-energy border. This compensation allows to use the continuous energy loss
approximation as the correct mean value of the propagated spectrum.

\section{EFFECT OF $\nu_\tau$ REGENERATION}
\label{sec:reg}

In this section, we present the results of a study of the effects of  $\nu_\tau$ and $\tau$ regeneration on the flux of emerging
$\tau$ leptons in different scenarios for the neutrino-nucleon
cross-section and for the $\tau$ energy losses. For this, we go back to the full propagation equations (\ref{eq:PhiNuTau}) and (\ref{eq:PhiTau1}), including all possible regeneration terms, except that we replace the stochastic energy loss term for the $\tau$ by the CEL expression as given in Eq. (\ref{eq:PhiTau2}), in concordance with the conclusions of Section \ref{sec:Taulosses}.  As stated in Section \ref{sec:tpt}, different calculations of these quantities exist in the literature. For the purpose of our analysis we have extracted a panel of {\em ad-hoc} parameterizations representing respectively a {\em low}, a {\em standard (std)} and a {\em high} value of either the $\tau$ energy loss parameter $\beta$ or neutrino-nucleon cross section $\sigma_{\nu N}$ (the explicit expressions are given in \cite{nousreg}). We solve the transport equations through a Monte Carlo simulation (which was also cross-checked against an iterative semi-analytical method: see the appendix of \cite{nousreg}) and compare the results obtained for different combinations of $\beta_i \otimes \sigma^{\nu N}_j$ ({\em i,j=low, std, high}).

\begin{figure}[!b]
    \centering
 \vspace*{-0.7cm}
    \includegraphics[width=7.5cm]{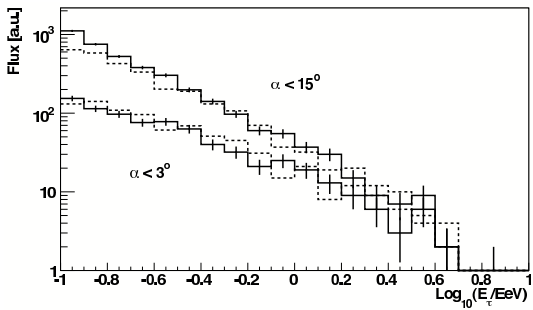}
\vspace*{-1cm}
    \caption{\footnotesize{Flux of emerging taus corresponding to incident neutrinos with
     an angle $\alpha$ up to 3$^{\circ}$ or 15$^{\circ}$ and $\Phi(E_\nu)\propto E_\nu^{-2}$ using the standard combination of parameterisations $\sigma_{std}\otimes\beta_{std}$.
    The regeneration is neglected in the dashed histograms and included in the solid ones.}}
    \label{fig:StdE2_15Reg}
\end{figure}

As a starting point, one should 
notice that the regeneration is a second order process which
requires the $\tau$ to be converted back into a $\nu_\tau$ before
loosing too much energy, and then this $\nu_\tau$ to undergo another
CC weak interaction to produce a $\tau$ again. Hence, one expects
that a higher energy loss or a lower neutrino-nucleon cross-section will reduce the effect of the regeneration, while, on the other hand, a higher $\nu N$
cross-section or a lower energy loss will enhance it. 

To verify this effect, we have first propagated through the Earth incident beams  of monoenergetic $\nu_\tau$'s injected at an angle $\alpha$ up to $3^\circ$, which roughly corresponds to the angular range of sensitivity of a surface detector such as the one of the Pierre Auger Observatory. We find that for a {\em standard} choice of parameterizations  $\beta_{std} \otimes \sigma^{\nu N}_{std}$ (see \cite{nousreg} for more detail), the effect of regeneration is negligible for $E_{\nu_\tau} < 10^9$ GeV, but not at higher energies where it can significantly increase the flux of emerging $\tau$'s (15\% more $\tau$'s for 30 EeV incident neutrinos).

\begin{figure}[!b]
    \centering
 \vspace*{-0.7cm}
    \includegraphics[width=7.5cm]{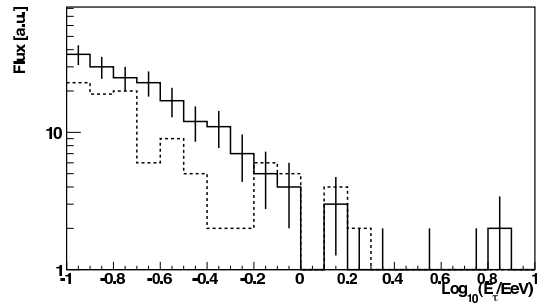}
\vspace*{-1cm}
    \caption{\footnotesize{Flux of emerging taus corresponding to incident neutrinos with
     angle up to 3$^{\circ}$ and $\Phi(E_\nu)\propto E_\nu^{-2}$ using the combination $\sigma_{high}\otimes\beta_{low}$.
    The regeneration is neglected in the dashed histogram and included in the solid one.}}
    \label{fig:HLE2_3degReg}
\end{figure}

As can be seen from Fig.\ref{fig:StdE2_15Reg}, this effect is washed out in the case of an incident neutrino flux $\Phi_{\nu_\tau} \propto E_\nu^{-2}$, for which the contribution of the highest-energy part of the neutrino spectrum is negligible; but it could affect the spectrum of emerging $\tau$'s if the neutrino flux were harder. Moreover, as shown in the same figure, even in the {\em standard} picture, the effect of the regeneration can be much more important for detectors which are sensitive to $\tau$'s emerging at larger angles $\alpha$: 30\% less $\tau$'s at 0.3 EeV for an incident neutrino flux  $\Phi(E_\nu)\propto E_\nu^{-2}$. This is typically the case for fluorescence telescopes such as the ones used by HiRes (and Auger), which detect the faint ultra-violet light emitted by nitrogen molecules that are excited as the shower traverses the atmosphere.

Going now to non-standard values of $\beta$ and $\sigma^{\nu N}$, the combination $\beta_{low} \otimes \sigma^{\nu N}_{high}$ is indeed the most sensitive to regeneration processes. In that case, neglecting the regeneration effects leads to a significant underestimation of the flux of emerging taus even for incident neutrinos with relatively low energies: about 70\% of the $\tau$'s are lost when injecting neutrinos with energy 0.3 EeV. This statement remains true even for incident neutrino fluxes $\Phi(E_\nu)\propto E_\nu^{-2}$ when integrating over a small angular range ($0^\circ \leq \alpha \leq 3^\circ$), as can be seen in Fig. \ref{fig:HLE2_3degReg}.

\section{CONCLUSIONS}
\label{sec:conc}

 In our studies\cite{nousEloss,nousreg}, we have questioned the validity of some approximations usually made when dealing with the propagation of UHE tau leptons and neutrinos through the Earth. 

We have first shown that the stochastic nature of the
radiative processes undergone by tau leptons at ultra-high
energy is indeed responsible for large fluctuations in the tau energy losses.
At the same time, however, these fluctuations are not so
large as to blur the picture with respect to the continuous energy
loss approximation, as far as the calculation concerns power-law injection spectra in a given, continuous energy range.

%This study was done on basis of one particular model~\cite{dutta1} which is rather pessimistic in the $y$ range where the stochastic effects can lead to hard losses, in the sense that it leads to a fairly high rate of interactions. Any other model leading to a lower rate would not change the picture. Moreover, any other model leading to comparable or slightly greater rate would have to present a significantly harder differential cross section in term of $y$ to challenge our conclusions.

We have then studied the mechanism of regeneration of the $\nu_\tau$ flux while crossing the Earth, to investigate its effect on the flux of emerging $\tau$ leptons. Assuming a detector with an energy threshold of 0.1 EeV and sensitive only to taus with an emerging angle below 3$^\circ$, the effect of regeneration is negligible for a flux of incident neutrinos $\Phi(E_\nu)\propto E_\nu^{-2}$ with the standard values of cross-sections and energy losses. But we have shown that this is not valid for other assumptions on the detector performance, neither for less standard values of the weak cross-section or tau energy losses.  Moreover, the error made in the computation of the flux of emerging taus will also increase with the hardness of the flux of incident neutrinos since the contribution of the regeneration effect increases with the energy of the incident neutrino.

The simplification of neglecting the regeneration is thus only safe for particular values of the physical properties playing a role on the propagation and specific detectors. It may lead to a significant underestimation of the flux of emerging $\tau$'s when looking at non-standard values of the weak cross-section
or tau energy losses. Therefore, it should be carefully treated and accounted for when studying the systematics due to the uncertainties on those properties or while using the Earth-skimming technique to test for instance higher weak cross-sections. Similarly, one should carefully
check the effect for the characteristics of the actual detector before neglecting the regeneration.


\begin{thebibliography}{99}
\bibitem{nurev} see e.g.  J.~K.~Becker,
  %``Status of neutrino astronomy,''
  J.\ Phys.\ Conf.\ Ser.\  {\bf 136} (2008) 022055   for a recent review.


\bibitem{athar}
  H.~Athar, M.~Jezabek and O.~Yasuda,
  %``Effects of neutrino mixing on high-energy cosmic neutrino flux,''
  Phys.\ Rev.\  D {\bf 62} (2000) 103007; D.V. Ahluwalia, Mod.\ Phys.\
  Lett.~{\bf A16}, 917 (2000).

\bibitem{halzen}  F.~Halzen and D.~Saltzberg,
  %``Tau neutrino appearance with a 1000-Megaparsec baseline,''
  Phys.\ Rev.\ Lett.\  {\bf 81} (1998) 4305.


\bibitem{nutau} A.~Letessier-Selvon, AIP Conf.Proc. {\bf 566} (2000) 157-171, arXiv:0009444 [astro-ph]; D.~Fargion, {\em Astrophys.\ J.}\ {\bf 570} (2002) 909 and references therein.

\bibitem{augernu}  J.~Abraham {\it et al.}  [The Pierre Auger Collaboration],
  %``Upper limit on the diffuse flux of UHE tau neutrinos from the Pierre Auger
  %Observatory,''
  Phys.\ Rev.\ Lett.\  {\bf 100}, 211101 (2008).

\bibitem{hiresnu} K. Martens [HiRes Collaboration], arXiv:0707.4417 [astro-ph].

\bibitem{nousEloss}  O.~B.~Bigas, O.~Deligny, K.~Payet and V.~Van Elewyck,
  %``Tau energy losses at ultra-high energy: continuous versus stochastic
  %treatment,''
  Phys.\ Rev.\  D {\bf 77} (2008) 103004.

\bibitem{nousreg}  O. B.~Bigas, O.~Deligny, K.~Payet and V.~Van Elewyck,
  %``UHE tau neutrino flux regeneration while skimming the Earth,''
  Phys.\ Rev.\  D {\bf 78} (2008) 063002.

\bibitem{stdelosses} L.~B.~Bezrukov and E.~V.~Bugaev,
  Yad.\ Fiz.\  {\bf 33} (1981) 1195; E.~V.~Bugaev and Yu.~V.~Shlepin,
  Phys.\ Rev.\  D {\bf 67} (2003) 034027; A. V. Butkevitch and S.P. Mikheyev,
  Zh. Eksp. Teor. Fiz. {\bf 122} (2002) 17;
  K.~S.~Kuzmin, K.~S.~Lokhtin and S.~I.~Sinegovsky,
  Int.\ J.\ Mod.\ Phys.\  A {\bf 20} (2005) 6956.

\bibitem{dutta1} S.~I.~Dutta et al.,
  Phys.\ Rev.\  D {\bf 63} (2001) 094020
  [arXiv:hep-ph/0012350].


\bibitem{PT} A.~A.~Petrukhin and D.~A.~Timashkov,
  %``Proton structure functions over the entire kinematical region,''
  Phys.\ Atom.\ Nucl.\  {\bf 67} (2004) 2216
  [Yad.\ Fiz.\  {\bf 67} (2004) 2241].

\bibitem{parente} N. Armesto, C. Merino, G. Parente and E. Zas,
  [arXiv:hep-ph/0709.4461].

\bibitem{sarkar}  L.~A.~Anchordoqui, A.~M.~Cooper-Sarkar, D.~Hooper and S.~Sarkar, %``Probing low-x QCD with ultra-high energy cosmic neutrinos at Auger,''
  Phys.\ Rev.\  D {\bf 74} (2006) 043008.

\bibitem{CGC}  L.~D.~McLerran and R.~Venugopalan,
  %``Green's functions in the color field of a large nucleus,''
  Phys.\ Rev.\  D {\bf 50} (1994) 2225.

\bibitem{han}  see T.~Han and D.~Hooper,
  %``The particle physics reach of high-energy neutrino astronomy,''
  New J.\ Phys.\  {\bf 6} (2004) 150 and references therein.

\bibitem{music} P.~Antonioli et al.,
  Astropart.\ Phys.\ {\bf 7} (1997) 357;  S.~I.~Dutta et al.,
  Phys.\ Rev.\  D {\bf 72} (2005) 013005
  [arXiv:hep-ph/0504208]. 
\end{thebibliography}
\end{document}